\documentclass[12pt]{article}
\usepackage{amsmath,amsfonts,amssymb}

\begin{document}
\thispagestyle{empty}

\def\theequation{\arabic{section}.\arabic{equation}}
\def\a{\alpha}
\def\b{\beta}
\def\g{\gamma}
\def\d{\delta}
\def\dd{\rm d}
\def\e{\epsilon}
\def\ve{\varepsilon}
\def\z{\zeta}
\def\B{\mbox{\bf B}}\def\cp{\mathbb {CP}^3}

\newcommand{\h}{\hspace{0.5cm}}

\begin{titlepage}
\renewcommand{\thefootnote}{\fnsymbol{footnote}}
\begin{center}
{\Large \bf Finite-size Effect of the Dyonic Giant Magnons in}
\vskip .5cm
{\Large \bf ${\cal N}=6$ super Chern-Simons Theory}
\end{center}
\vskip 1.2cm \centerline{\bf Changrim  Ahn and P. Bozhilov
\footnote{On leave from Institute for Nuclear Research and Nuclear
Energy, Bulgarian Academy of Sciences, Bulgaria.}}

\vskip 10mm

\centerline{\sl Department of Physics} \centerline{\sl Ewha Womans
University} \centerline{\sl DaeHyun 11-1, Seoul 120-750, S. Korea}
\vspace*{0.6cm} \centerline{\tt ahn@ewha.ac.kr,
bozhilov@inrne.bas.bg}

\vskip 20mm

\baselineskip 18pt

\begin{center}
{\bf Abstract}
\end{center}
\h We consider finite-size effects for the dyonic giant magnon of the type IIA string theory
on $AdS_4\times \mathbb{CP}^3$ by applying L\"uscher $\mu$-term formula which is derived
from a recently proposed $S$-matrix for the ${\cal N}=6$ super Chern-Simons theory.
We compute explicitly the effect for the case of a symmetric configuration
where the two external bound states, each of $A$ and $B$ particles, have the same
momentum $p$ and spin $J_2$.
We compare this with the classical string theory result which
we computed by reducing it to the Neumann-Rosochatius system.
The two results match perfectly.

\end{titlepage}
\newpage
\baselineskip 18pt

\def\nn{\nonumber}
\def\tr{{\rm tr}\,}
\def\p{\partial}
\newcommand{\non}{\nonumber}
\newcommand{\bea}{\begin{eqnarray}}
\newcommand{\eea}{\end{eqnarray}}
\newcommand{\bde}{{\bf e}}
\renewcommand{\thefootnote}{\fnsymbol{footnote}}
\newcommand{\be}{\begin{eqnarray}}
\newcommand{\ee}{\end{eqnarray}}

\vskip 0cm

\renewcommand{\thefootnote}{\arabic{footnote}}
\setcounter{footnote}{0}

\setcounter{equation}{0}
\section{Introduction}

The AdS/CFT correspondence between type IIB string
theory on $AdS_5\times S^5$ and ${\cal N}=4$ super Yang-Mills (SYM) theory
\cite{Maldacena,GKP,Witten} led to many exciting developments and to
understanding non-perturbative sturctures of the string and gauge theories.
Recently, an exciting possibility that the same type of duality does exist in
three-dimensional gauge theory has been discovered.
The promising candidate for the three-dimensional conformal field theory
is ${\cal N}=6$ super Chern-Simons (CS) theory
with $SU(N)\times SU(N)$ gauge symmetry and level $k$.  This model,
which was first proposed by Aharony, Bergman, Jafferis, and Maldacena
\cite{ABJM}, is believed to be dual to M-theory on $AdS_4\times
S^7/Z_k$.  Furthermore, in the planar limit of $N, k\to\infty$ with a
fixed value of 't Hooft coupling $\lambda=N/k$, the ${\cal N}=6$ CS is
believed to be dual to type IIA superstring theory on $AdS_4\times
\mathbb{CP}^3$.  This model contains two sets of scalar fields transforming in
bifundamental representations of $SU(N)\times SU(N)$ along with
respective superpartner fermions and non-dynamic CS gauge
fields.

The integrability of the planar ${\cal N}=6$ CS theory, first discovered by
Minahan and Zarembo \cite{MZ} in the leading two-loop-order
perturbative computation, is conjectured to exist in all-loop orders
and corresponding all-loop Bethe ansatz equations were conjectured
by Gromov and Vieira \cite{GV} based on the perturbative result \cite{MZ}
and the classical integrability in the large-coupling limit
discovered in \cite{AruFro,Stefj,GVi}.
Recently, three groups \cite{MR, AAB, Kr} computed the one-loop
correction to the energy of a folded spinning string, and seemed to
find disagreement with the prediction of the all-loop BAEs.  This
controversy may be resolved by a non-zero one-loop correction in the
central interpolating function $h(\lambda)$ as suggested recently in
\cite{MRT}.  (See also \cite{GM}.)

On the other hand, based on the spectrum and symmetries of the model
\cite{MZ, NT, GGY,GHO}, Ahn and Nepomechie proposed an $S$-matrix \cite{AN}
which reproduces the all-loop BAEs.
The $S$-matrix has played an important role in AdS/CFT as noticed early in
\cite{Staudacher} and gets more so because it provides only way of computing
finite-size effect exactly.
For example, one can not reproduce this from the all-loop BAEs.
Therefore, the finite-size effect can be a stringent check of the $S$-matrix
in integrable models if one can compare it with an
independent result.
In the AdS/CFT correspondence, there are alternative but approximate ways of computing finite-size
effects in semi-classical ways \cite{AFZf} such as algebraic curve method
\cite{GSV} or Neumann-Rosochatius method \cite{AFRT,ART,KRT06}.
For the ${\cal N}=6$ CS theory, both methods have been implemented to compute the effect
for a giant magnon (GM) which moves symmetrically in $SU(2)\times SU(2)$ subspace of
$\mathbb{CP}^3$ \cite{APGHO}.
See \cite{Shender,GHOS,AB,LPP,BT}
for subsequent developments on the finite-size effects of the ${\rm AdS}_4/{\rm CFT}_3$
from the string/membrane side.

The formalism to derive the finite-size effect from the $S$-matrix
is the L\"uscher correction.
This computes a shift in the energy due to
the finite-size of spatial length from the $S$-matrix
for all values of the `t Hooft coupling constant.
This method has been successfully applied to the AdS/CFT duality in the ${\cal N}=4$
SYM theory \cite{AJK,JL,HatSuzi,BajJan,HatSuzii}.
Recently two groups computed the finite-size
corrections to the dispersion relation of GMs \cite{HM} from the
${\cal N}=6$ CS theory side \cite{BF,LS}.
They showed that the results are consistent with the classical string theory,
which strongly supports the validity
of the $S$-matrix proposed in \cite{AN}.
Along this line of investigation,
another interesting configuration is the classical string state with two angular
momenta, usually called `two spin' solutions.
The authors have computed the finite-size effect for the dyonic giant
magnon (DGM) \cite{ND} in the classical limit by the Neumann-Rosochatius method \cite{ABR}.
It is further extended recently to a single DGM solution \cite{Ryang}.
It is important to check this since the DGM maintains the BPS saturated form of
the dispersion relation even in the classical limit.
Therefore, it can check the finite-size effect in a most intact form.

The purpose of this note is to compute the finite-size effect of the DGM from the
L\"uscher formula and compare it with the previous result \cite{ABR}.
In sect.2, we briefly review our derivation of the DGM in the $SU(2)\times SU(2)$ subspace of
$\mathbb{CP}^3$ with $U(1)$ fibre dynamics and computation of the finite-size effect \cite{ABR}.
We generalize the L\"uscher formula for multi-DGM particles in sect.3.
We also derive the $S$-matrix elements between an elementary magnon
and its bound state in sect.3 which will be used in the L\"uscher formula.
In the classical limit, we confirm that the L\"uscher correction exactly matches with
the classical string theory result.
We conclude the paper with some remarks in sect.4.

\setcounter{equation}{0}
\section{Classical string analysis}
Let us consider a classical string moving in $R_t\times\cp$. Using
the complex coordinates \bea\nn z=y^0+iy^4,\h w_1=x^1+ix^2,\h
w_2=x^3+ix^4,\h w_3=x^5+ix^6,\h w_4=x^7+ix^8,\eea we embed the
string as follows \cite{ABR} \bea\nn
&&z=Z(\tau,\sigma)=\frac{R}{2}e^{it(\tau,\sigma)},\h
w_a=W_a(\tau,\sigma)=R r_a(\tau,\sigma)e^{i\varphi_a(\tau,\sigma)}.
\eea Here $t$ is the $AdS$ time. These complex coordinates should
satisfy \bea \nn &&\sum_{a=1}^{4}W_a\bar{W}_a=R^2,\h
\sum_{a=1}^{4}\left(W_a\p_m\bar{W}_a-\bar{W}_a\p_m W_a\right)=0,
\eea or \bea \sum_{a=1}^{4}r_a^2=1,\h
\sum_{a=1}^{4}r_a^2\p_m\varphi_a=0,\quad m=0,1. \label{cp3cond} \eea

\subsection{NR Reduction}

In order to reduce the string dynamics on $R_t\times\cp$ to the NR
integrable system, we use the ansatz \cite{AFRT,ART,KRT06}
\bea\label{NRA} &&t(\tau,\sigma)=\kappa\tau,\h
r_a(\tau,\sigma)=r_a(\xi),\h
\varphi_a(\tau,\sigma)=\omega_a\tau+f_a(\xi),\\ \nn
&&\xi=\alpha\sigma+\beta\tau,\h \kappa, \omega_a, \alpha, \beta={\rm
constants}.\eea It can be shown \cite{ABR} that after integration of
the equations of motion for $f_a$, which gives \bea\label{fafi}
f'_a=\frac{1}{\alpha^2-\beta^2}
\left(\frac{C_a}{r_a^2}+\beta\omega_a\right),\h C_a=constants, \eea
one ends up with the following effective Lagrangian for the
coordinates $r_a$ \bea \label{NRL} L_{NR}&=&(\alpha^2-\beta^2)
\sum_{a=1}^{4}\left[r^{'2}_a-\frac{1}{(\alpha^2-\beta^2)^2}
\left(\frac{C_a^2}{r_a^2} + \alpha^2\omega_a^2r_a^2\right)\right]
\\ \nn &-&\Lambda\left(\sum_{a=1}^{4}r_a^2-1\right) .\eea
This is the Lagrangian for the NR integrable system \cite{KRT06}. In
addition, the $\cp$ embedding conditions in (\ref{cp3cond}) lead to
\bea \sum_{a=1}^{4}\omega_a r_a^2=0,\h
\sum_{a=1}^{4}C_a=0.\label{aec} \eea

The Virasoro constraints give the conserved Hamiltonian $H_{NR}$ and
a relation between the embedding parameters and the arbitrary
constants $C_a$: \bea\label{HNR} &&H_{NR}=(\alpha^2-\beta^2)
\sum_{a=1}^{4}\left[r_a'^2+\frac{1}{(\alpha^2-\beta^2)^2}
\left(\frac{C_a^2}{r_a^2} + \alpha^2\omega_a^2r_a^2\right)\right]
=\frac{\alpha^2+\beta^2}{\alpha^2-\beta^2}\frac{\kappa^2}{4},
\\ \label{01R} &&\sum_{a=1}^{4}C_a\omega_a + \beta(\kappa/2)^2=0.\eea
For closed strings, $r_a$ and $f_a$ satisfy the following
periodicity conditions \bea r_a(\xi+2\pi\alpha)=r_a(\xi),\h
f_a(\xi+2\pi\alpha)=f_a(\xi)+2\pi n_a,\label{pbc} \eea where $n_a$
are integer winding numbers.

The conserved charges can be defined by
\bea\nn E_s=-\int
d\sigma\frac{\p\mathcal{L}}{\p(\p_0 t)},\h J_a=\int
d\sigma\frac{\p\mathcal{L}}{\p(\p_0\varphi_a)},\h a=1,2,3,4,\eea
where $\mathcal{L}$ is the Polyakov string Lagrangian taken in
conformal gauge. Using the ansatz (\ref{NRA}) and (\ref{fafi}), we
can find \bea\label{cqs} E_s=
\frac{\kappa\sqrt{2\lambda}}{2\alpha}\int d\xi,\h J_a=
\frac{2\sqrt{2\lambda}}{\alpha^2-\beta^2}\int d\xi
\left(\frac{\beta}{\alpha}C_a+\alpha\omega_a r_a^2\right).\eea In
view of (\ref{aec}), one obtains \cite{GHO}  \bea\label{Jc}
\sum_{a=1}^{4}J_a=0.\eea

\subsection{Dyonic Giant Magnon Solution}

We are interested in finding string configurations corresponding to
the following particular solution of (\ref{aec}) \bea\nn
r_1=r_3=\frac{1}{\sqrt{2}}\sin\theta,\h
r_2=r_4=\frac{1}{\sqrt{2}}\cos\theta,\h \omega_1=-\omega_3,\h
\omega_2=-\omega_4.\eea The two frequencies $\omega_1, \omega_2$ are
independent and lead to strings moving in $\cp$ with two angular
momenta. The special case $\omega_2=0$ corresponds to the solutions
obtained in \cite{GHO,GHOS}. From the NR Hamiltonian (\ref{HNR}) one
finds \bea\nn \theta'^2(\xi)=\frac{1}{(\alpha^2-\beta^2)^2}
\left[\frac{\kappa^2}{4}(\alpha^2+\beta^2) -
2\left(\frac{C_1^2+C_3^2}{\sin^2{\theta}} +
\frac{C_2^2+C_4^2}{\cos^2{\theta}}\right)-
\alpha^2\left(\omega_1^2\sin^2{\theta}
+\omega_2^2\cos^2{\theta}\right)\right]. \eea We further restrict
ourselves to $C_2=C_4=0$ to search for GM string configurations.
Eqs. (\ref{aec}) and (\ref{01R}) give \bea\nn
C_1=-C_3=-\frac{\beta\kappa^2}{8\omega_1}. \eea In this case, the
above equation for $\theta'$ can be rewritten in the form
\bea\label{tS3eq}
(\cos\theta)'=\mp\frac{\alpha\sqrt{\omega_1^2-\omega_2^2}}{\alpha^2-\beta^2}
\sqrt{(z_+^2-\cos^2\theta)(\cos^2\theta-z_-^2)},\eea where \bea\nn
&&z^2_\pm=\frac{1}{2(1-\frac{\omega_2^2}{\omega_1^2})}
\left\{y_1+y_2-\frac{\omega_2^2}{\omega_1^2}
\pm\sqrt{(y_1-y_2)^2-\left[2\left(y_1+y_2-2y_1
y_2\right)-\frac{\omega_2^2}{\omega_1^2}\right]
\frac{\omega_2^2}{\omega_1^2}}\right\}, \\ \nn
&&y_1=1-\frac{\kappa^2}{4\omega_1^2},\h
y_2=1-\frac{\beta^2}{\alpha^2}\frac{\kappa^2}{4\omega_1^2}.\eea The
solution of (\ref{tS3eq}) is given by \bea\label{tS3sol}
\cos\theta=z_+ dn\left(C\xi|m\right),\h
C=\mp\frac{\alpha\sqrt{\omega_1^2-\omega_2^2}}{\alpha^2-\beta^2}
z_+,\h m\equiv 1-z^2_-/z^2_+ ,\eea where $dn\left(C\xi|m\right)$ is
one of the elliptic functions.

To find the full string solution, we also need to obtain the
explicit expressions for the functions $f_a$ from (\ref{fafi})
\bea\nn f_a=\frac{1}{\alpha^2-\beta^2}\int
d\xi\left(\frac{C_a}{r_a^2}+\beta\omega_a\right).\eea Using the
solution (\ref{tS3sol}) for $\theta(\xi)$, we can find \bea\nn
&&f_1=-f_3=\frac{\beta/\alpha}{z_+\sqrt{1-\omega_2^2/\omega_1^2}}
\left[C\xi -
\frac{2(\kappa/2)^2/\omega_1^2}{1-z^2_+}\Pi\left(am(C\xi),-\frac{z^2_+
-z^2_-}{1-z^2_+}|m\right)\right],\\ \nn
&&f_2=-f_4=\frac{\beta\omega_2}{\alpha^2-\beta^2}\xi.\eea Here,
$\Pi$ is the elliptic integral of the third kind. As a consequence,
the string solution can be written as \bea \nn
&&W_1=\frac{R}{\sqrt{2}}\sqrt{1-z_+^2dn^2\left(C\xi|m\right)}\
e^{i(\omega_1\tau+f_1)},\\ \label{fss} &&W_2=\frac{R}{\sqrt{2}}z_+
dn\left(C\xi|m\right)\ e^{i(\omega_2\tau+f_2)},\\ \nn
&&W_3=\frac{R}{\sqrt{2}}\sqrt{1-z_+^2dn^2\left(C\xi|m\right)}\
e^{-i(\omega_1\tau+f_1)},\\ \nn &&W_4=\frac{R}{\sqrt{2}}z_+
dn\left(C\xi|m\right)\ e^{-i(\omega_2\tau+f_2)} .\eea

The geometric meaning of the explicit solution (\ref{fss}) is as
follows. Each pairs of complex coordinates, $(W_1,W_2)$ and
$(W_3,W_4)$, describe a spiky solutions in $S^2$ sphere geometry but
with dynamics at opposite points in the $U(1)$ fiber. The two phases
in $(W_1,W_2)$ are exactly opposite to those of $(W_3,W_4)$ which,
together with the dynamics in $U(1)$, ensures the vanishing of the
total momentum. This behavior has been also noticed for strings in
$R_t\times S^2\times S^2$ in \cite{GHO}.

The GM in infinite volume can be obtained by taking $z_-\to 0$. In
this limit, the solution for $\theta$ reduces to \bea\nn
\cos\theta=\frac{\sin\frac{p}{2}}{\cosh(C\xi)}, \eea where the
constant $z_{+}\equiv\sin p/2$ is given by \bea\nn
z^2_{+}=\frac{y_2-\omega_2^2/\omega_1^2}{1-\omega_2^2/\omega_1^2}.
\eea One spin solution corresponds $\omega_2=0$.
Inserting this into (\ref{cqs}), one can find the energy-charge
dispersion relation. For the {\it single} DGM, the energy and
angular momentum $J_1$ become infinite but their difference remains
finite:
\bea E_s-J_1=\sqrt{\frac{J_2^2}{4}+2\lambda\sin^2\frac{p}{2}}.
\label{infdis} \eea

\subsection{Finite-size Effects}

Using the most general solutions (\ref{fss}), we can calculate the
finite-size corrections to the energy-charge relation (\ref{infdis})
in the limit when the string energy $E_s\to\infty$. Here we consider
the case of $\alpha^2>\beta^2$ only since it corresponds to the GM
case. We obtain from (\ref{cqs}) the following expressions for the
conserved string energy $E_s$ and the angular momenta $J_a$ \bea\nn
&&\mathcal{E} =\frac{2\kappa(1-\beta^2/\alpha^2)} {\omega_1
z_+\sqrt{1-\omega_2^2/\omega_1^2}}\mathbf{K}
\left(1-z^2_-/z^2_+\right), \\ \label{cqsGM} &&\mathcal{J}_1=
\frac{2 z_+}{\sqrt{1-\omega_2^2/\omega_1^2}} \left[
\frac{1-\beta^2(\kappa/2)^2/\alpha^2\omega_1^2}{z^2_+}\mathbf{K}
\left(1-z^2_-/z^2_+\right)-\mathbf{E}
\left(1-z^2_-/z^2_+\right)\right], \\ \nn &&\mathcal{J}_2= \frac{2
z_+ \omega_2/\omega_1 }{\sqrt{1-\omega_2^2/\omega_1^2}}\mathbf{E}
\left(1-z^2_-/z^2_+\right),\h \mathcal{J}_3=-\mathcal{J}_1,\h
\mathcal{J}_4=-\mathcal{J}_2.\eea As a result, the condition
(\ref{Jc}) is identically satisfied. Here, we introduced the
notations \bea\label{not} \mathcal{E}=\frac{E_s}{\sqrt{2\lambda}}
,\h \mathcal{J}_a=\frac{J_a}{\sqrt{2\lambda}}.\eea The computation
of $\Delta\varphi_1$ gives \bea\label{pws} p\equiv\Delta\varphi_1
&=& 2\int_{\theta_{min}}^{\theta_{max}}\frac{d \theta}{\theta'}f'_1=
\\ \nn &-&\frac{2\beta/\alpha}{z_+\sqrt{1-\omega_2^2/\omega_1^2}}
\left[\frac{(\kappa/2)^2/\omega_1^2}{1-z^2_+}\Pi\left(-\frac{z^2_+ -
z^2_-}{1-z^2_+}\bigg\vert 1-z^2_-/z^2_+\right) -\mathbf{K}
\left(1-z^2_-/z^2_+\right)\right].\eea In the above expressions,
$\mathbf{K}(m)$, $\mathbf{E}(m)$ and $\Pi(n|m)$ are the complete
elliptic integrals.

Expanding the elliptic integrals, we obtain
\bea
E-J_1&=& 2\sqrt{\frac{J_2^2}{4}+2\lambda\sin^2\frac{p}{2}}\label{stringDGM}\\
&-&\frac{32\lambda\sin^4\frac{p}{2}}
{\sqrt{J_2^2+8\lambda\sin^2\frac{p}{2}}}\exp\left[-\frac{2\sin^2\frac{p}{2}\left(J_1 +
\sqrt{J_2^2+8\lambda\sin^2\frac{p}{2}}\right)
\sqrt{J_2^2+8\lambda\sin^2\frac{p}{2}}}{J_2^2+8\lambda\sin^4\frac{p}{2}}
\right].\nn
\eea
This also gives a finite-size effect for ordinary GM \cite{AFZf} by taking $J_2\to 0$
\bea
E-J_1= 2\sqrt{2\lambda}\sin\frac{p}{2}-16\sqrt{\frac{\lambda}{2}}\sin^3\frac{p}{2}
\exp\left[-\frac{J_1}{\sqrt{2\lambda}\sin\frac{p}{2}}-2\right].
\eea

\setcounter{equation}{0}
\section{Finite-size effects from the $S$-matrix}

The ${\cal N}=6$ CS theory has two sets of
excitations, namely $A$-particles and $B$-particles, each of which
form a four-dimensional representation of $SU(2|2)$ \cite{GGY,AN}.
We propose an $S$-matrix with the following structure:
\be
S^{AA}(p_1,p_2)&=&S^{BB}(p_1,p_2)=S_0(p_1,p_2){\widehat S}(p_1,p_2)
\,, \non \\
S^{AB}(p_1,p_2)&=&S^{BA}(p_1,p_2)={\tilde S}_0(p_1,p_2){\widehat
S}(p_1,p_2) \,, \non
\ee
where $\widehat S$ is the matrix part determined by the $SU(2|2)$
symmetry, and is essentially the same as that found for ${\cal N}=4$ YM in \cite{Be,AFZ}.
An important difference arises in the dressing phases $S_0, {\tilde
S}_0$ due to the fact that the
$A$- and $B$-particles are related by complex conjugation.

\subsection{L\"uscher $\mu$-term Formula}

Here we want to generalize multi-particle L\"uscher formula \cite{BajJan,HatSuzii} to the
case of the bound states.
Consider $M_A$ number of A-type DGMs, $\vert Q_1,\ldots Q_{M_A}\rangle$, and
$M_B$ number of B-type DGMs, $\vert {\tilde Q}_1,\ldots {\tilde Q}_{M_B}\rangle$.
We use $\alpha_k$ for the $SU(2|2)$ quantum numbers carried by the DGMs and
$C_k$ for $A$ or $B$, the two types of particles.
Then we propose the multi-particle L\"uscher formula for generic DGM states as follows:
\be
&&\delta E_{\mu}=-i\sum_{b=1}^4\left\{\sum_{l=1}^{M_A}(-1)^{F_b}
\left(1-\frac{\epsilon'_{Q_{l}}(p_l)}{\epsilon'_{1}({\tilde q}^*)}\right)
e^{-i{\tilde q}^* L}\left[\mathop {\rm Res} \limits_{q^*={\tilde q}^*}
{S^{AA}}^{b\alpha_{l}}_{b\alpha_{l}}(q^*,p_l)\right]
\prod_{k\neq l}^{M_A+M_B}{S^{AC_k}}^{b\alpha_{k}}_{b\alpha_{k}}(q^*,p_k)\right.\non\\
&&+\left.\sum_{l=1}^{M_B}(-1)^{F_b} \left(1-\frac{\epsilon'_{{\tilde
Q}_{l}}(p_l)}{\epsilon'_{1}({\tilde q}^*)}\right) e^{-i{\tilde q}^*
L}\left[\mathop {\rm Res} \limits_{q^*={\tilde q}^*}
{S^{BB}}^{b\alpha_{l}}_{b\alpha_{l}}(q^*,p_l)\right] \prod_{k\neq
l}^{M_A+M_B}{S^{BC_k}}^{b\alpha_{k}}_{b\alpha_{k}}(q^*,p_k)\right\}.
\label{luscher} \ee Here, the energy dispersion relation for the DGM
is given by \be
\epsilon_Q(p)=\sqrt{\frac{Q^2}{4}+4g^2\sin^2\frac{p}{2}}.
\label{disper} \ee
Here the coupling constant $g=h(\lambda)$ is still unknown function of $\lambda$ which behaves
as $h(\lambda) \sim \lambda$ for small $\lambda$, and
$h(\lambda) \sim \sqrt{\lambda/2}$ for large $\lambda$.

\subsection{$S$-matrix elements for the Dyonic GM}
The $S$-matrix elements for the DGM are in general complicated.
However, we can consider a simplest case of the DGMs composed of only A-type $\phi_1$'s which
are the first bosonic particle in the fundamental representation of $SU(2|2)$.
It is obvious that these bound states do exist since the elementary $S$-matrix element
$S{^{AA}}^{11}_{11}$ does have a pole.
The same holds for the B-type DGMs.
However, the hybrid type DGMs are not possible because the $S^{AB}$ $S$-matrix does not have
any bound-state pole.

The L\"uscher correction needs only those $S$-matrix elements
which have the same incoming and outgoing
$SU(2|2)$ quantum numbers after scattering with a virtual particle.
In particular, we can easily compute the matrix elements between an elementary
magnon and a the bound-state made of only $\phi_1$'s ($Q$ of them) denoted by ${\mathbf 1}_Q$
\cite{HatSuzi}
\be
{S^{AA}}^{b\mathbf{1}_Q }_{b {\mathbf 1}_Q}(y,{X^{(Q)}})=\prod_{k=1}^Q
{S^{AA}}^{b 1}_{b 1}(y,x_k)
=\prod_{k=1}^Q\left[
\frac{1-\frac{1}{y^+x^-_k}}{1-\frac{1}{y^-x^+_k}}\sigma_{\rm BES}(y,x_k)
{\tilde a}_b(y,x_k)\right],
\ee
where ${\tilde a}_b$ are given by \cite{Be,AFZ}
\be
{\tilde a}_1(y,x)&=&a_1(y,x),\quad {\tilde a}_2(y,x)=a_1(y,x)+a_2(y,x),
\quad {\tilde a}_3(y,x)={\tilde a}_4(y,x)=a_6(y,x)\non\\
a_1(y,x)&=&\frac{x^--y^+}{x^+-y^-}\frac{\eta(x)\eta(y)}{{\tilde\eta}(x){\tilde\eta}(y)}\non\\
a_2(y,x)&=&\frac{(y^--y^+)(x^--x^+)(x^--y^+)}{(y^--x^+)(x^-y^--x^+y^+)}
\frac{\eta(x)\eta(y)}{{\tilde\eta}(x){\tilde\eta}(y)}\non\\
a_6(y,x)&=&\frac{y^+-x^+}{y^--x^+}\frac{\eta(y)}{{\tilde\eta}(y)}.\non
\ee

As noticed in \cite{HatSuzi}, $a_2/a_1$ and $a_6/a_1$ are negligible ${\cal O}(1/g)$
corrections in the classical limit $g>>1$.
Therefore, the $S$-matrix with $b=1$ is a most important factor for our computation
which can be written as
\be
{S^{AA}}^{1 \mathbf{1}_{Q}}_{1 \mathbf{1}_{Q}}(y,{X^{(Q)}})&=&\sigma_{\rm BES}(y,X^{(Q)})
\prod_{k=1}^Q\left[\frac{1-\frac{1}{y^+x^-_k}}{1-\frac{1}{y^-x^+_k}}\cdot
\frac{x^-_k-y^+}{x^+_k-y^-}\frac{\eta(x_k)\eta(y)}{{\tilde\eta}(x_k){\tilde\eta}(y)}\right]\non\\
&=&\sigma_{\rm BES}(y,X^{(Q)})S_{\rm BDS}(y,X^{(Q)})
\frac{\eta(X^{(Q)})}{{\tilde\eta}(X^{(Q)})}\left(\frac{\eta(y)}{{\tilde\eta}(y)}\right)^Q,
\label{saa}\\
{S^{AB}}^{1 \mathbf{1}_{Q}}_{1
\mathbf{1}_{Q}}(y,{X^{(Q)}})&=&\sigma_{\rm BES}(y,X^{(Q)})
\frac{\eta(X^{(Q)})}{{\tilde\eta}(X^{(Q)})}\left(\frac{\eta(y)}{{\tilde\eta}(y)}\right)^Q,
\label{sab}
\ee
where the BDS  $S$-matrix is defined by
\be
S_{\rm BDS}(y,x)\equiv\frac{1-\frac{1}{y^+x^-}}{1-\frac{1}{y^-x^+}}\cdot
\frac{x^--y^+}{x^+-y^-}.
\ee
The spectral parameter $X^{(Q)}$ for the DGM is defined by
\be
{X^{(Q)}}^{\pm}=\frac{e^{\pm i p/2}}{4g\sin\frac{p}{2}}\left(Q
+\sqrt{Q^2+16g^2\sin^2\frac{p}{2}}\right)\equiv e^{(\theta\pm i
p)/2},
\ee
where we introduce $\theta$ defined by \be
\sinh\frac{\theta}{2}&\equiv&\frac{Q}{4g\sin\frac{p}{2}}. \ee
The frame factors $\eta$ and ${\tilde\eta}$ are given by \cite{AFZ}
\be
\frac{\eta(x_1)}{{\tilde\eta}(x_1)}=\frac{\eta(x_2)}{{\tilde\eta}(x_2)}=1
\ee
for the spin-chain frame and
\be
\frac{\eta(x_1)}{{\tilde\eta}(x_1)}=\sqrt{\frac{x^+_2}{x^-_2}},\quad
\frac{\eta(x_2)}{{\tilde\eta}(x_2)}=\sqrt{\frac{x^-_1}{x^+_1}}
\ee
for the string frame.

\subsection{Symmetric DGM state}
The classical two spins solution described in sect.2 is a symmetric DGM configuration for both of
$S^2$ subspaces.
Corresponding L\"uscher formula is given by Eq.(\ref{luscher}) with $M_A=M_B=1$,
which can be much simplified as
\be
\delta E_{\mu}&=&-i\sum_{b=1}^4 (-1)^{F_b}e^{-i{\tilde q}^* L}\left\{
\left(1-\frac{\epsilon'_{Q}(p_1)}{\epsilon'_{1}({\tilde q}^*)}\right)
\left[\mathop {\rm Res} \limits_{q^*={\tilde q}^*}
{S^{AA}}^{b\mathbf{1}_Q}_{b\mathbf{1}_Q}(q^*,p_1)\right]
{S^{AB}}^{b\mathbf{1}_{\tilde Q}}_{b\mathbf{1}_{\tilde Q}}(q^*,p_2)\right.\non\\
&+&\left.
\left(1-\frac{\epsilon'_{{\tilde Q}}(p_2)}{\epsilon'_{1}({\tilde q}^*)}\right)
\left[\mathop {\rm Res} \limits_{q^*={\tilde q}^*}
{S^{AA}}^{b\mathbf{1}_{\tilde Q}}_{b\mathbf{1}_{\tilde Q}}(q^*,p_2)\right]
{S^{AB}}^{b\mathbf{1}_Q}_{b\mathbf{1}_Q}(q^*,p_1)\right\}.
\label{luscher1}
\ee

As mentioned earlier, only the two cases of $b=1,2$ contributes equally in
the sum of Eq.(\ref{luscher1}) since these elements contain $a_1$.
Instead of the summation, we can multiply a factor 2 for the case of $b=1$.
In that case, we can compute easily each term using the $S$-matrix elements (\ref{saa})
and (\ref{sab}).
Furthermore, we restrict ourselves for the case where the two DGMs are symmetric in both
spheres, namely, $p_1=p_2$ and $Q={\tilde Q}$.
This leads to
\be
\delta E_{\mu}=-4i e^{-i{\tilde q}^* L}
\left(1-\frac{\epsilon'_{Q}(p)}{\epsilon'_{1}({\tilde q}^*)}\right)
\left[\mathop {\rm Res} \limits_{q^*={\tilde q}^*}
{S^{AA}}^{1\mathbf{1}_Q}_{1\mathbf{1}_Q}(q^*,p)\right]
{S^{AB}}^{1\mathbf{1}_Q}_{1\mathbf{1}_Q}(q^*,p).
\label{luscher2}
\ee

Explicit computations of each factor in (\ref{luscher2}) are exactly
the same as those in \cite{HatSuzi}.
There are two types of poles of $S_{\rm BDS}(y,X^{(Q)})$.
The $s$-channel pole which describe $(Q+1)$-DGM arises at $y^-={X^{(Q)}}^+$ while
the $t$-channel pole for $(Q-1)$-DGM (for $Q\ge 2$) at $y^+={X^{(Q)}}^+$.
We consider the $s$-channel pole first.
Using the location of the pole, we can find
\be
{\tilde q}^*=-\frac{i}{2g\sin\left(\frac{p-i\theta}{2}\right)}\quad \to\quad
e^{-i{\tilde q}^* L}\approx\exp\left[
-\frac{L}{2g\sin\left(\frac{p-i\theta}{2}\right)}\right].
\ee
From Eq.(\ref{disper}), one can also obtain
\be
1-\frac{\epsilon'_{Q}(p)}{\epsilon'_{1}({\tilde q}^*)}\approx
\frac{\sin\frac{p}{2}\sin\frac{p-i\theta}{2}}{\cosh\frac{\theta}{2}}.
\ee
Furthermore, one can notice from Eqs.(\ref{saa}) and (\ref{sab})
\be
\left[\mathop {\rm Res} \limits_{q^*={\tilde q}^*}
{S^{AA}}^{1\mathbf{1}_Q}_{1\mathbf{1}_Q}(q^*,p)\right]
{S^{AB}}^{1\mathbf{1}_Q}_{1\mathbf{1}_Q}(q^*,p)=
\mathop {\rm Res} \limits_{q^*={\tilde q}^*} {S_{\rm SYM}}^{1\mathbf{1}_Q}_{1\mathbf{1}_Q}(q^*,p)
\ee
where $S_{\rm SYM}$ is the $S$-matrix of the ${\cal N}=4$ SYM theory.
Explicit evaluation of the residue term becomes in the leading order
\be
-\frac{8ig e^{-ip}\sin^2\frac{p}{2}}{\sin\frac{p-i\theta}{2}}
\exp\left[-\frac{2e^{-\theta/2}\sin\frac{p}{2}}{\sin\frac{p-i\theta}{2}}\right]
\left(\frac{\eta(X^{(Q)})}{{\tilde\eta}(X^{(Q)})}\right)^2
\left(\frac{\eta(y)}{{\tilde\eta}(y)}\right)^{2Q}.
\ee
Combining all these together, we get
\be
\delta E_{\mu}&=&-\frac{8g e^{-ip}\sin^3\frac{p}{2}}{\cosh\frac{\theta}{2}}
\exp\left[-\frac{2e^{-\theta/2}\sin\frac{p}{2}}{\sin\frac{p-i\theta}{2}}
-\frac{L}{2g\sin\left(\frac{p-i\theta}{2}\right)}\right]
\left(\frac{\eta(X^{(Q)})}{{\tilde\eta}(X^{(Q)})}\right)^2
\left(\frac{\eta(y)}{{\tilde\eta}(y)}\right)^{2Q}\non\\
&=&-\frac{32g\sin^3\frac{p}{2}\ e^{i\alpha}}{\cosh\frac{\theta}{2}}\exp\left[
-\frac{2\sin^2\frac{p}{2}\cosh^2\frac{\theta}{2}}{\sin^2\frac{p}{2}+\sinh^2\frac{\theta}{2}}
\left(\frac{L-Q}{2g\sin\frac{p}{2}\cosh\frac{\theta}{2}}+1\right)\right]\\
&=&-\frac{32g^2\sin^4\frac{p}{2}\ e^{i\alpha}}{\sqrt{Q^2+16g^2\sin^2\frac{p}{2}}}
\exp\left[-\frac{2\sin^2\frac{p}{2}\left(L+\sqrt{Q^2+16g^2\sin^2\frac{p}{2}}\right)
\sqrt{Q^2+16g^2\sin^2\frac{p}{2}}}{Q^2+16g^2\sin^4\frac{p}{2}}\right].\non
\ee
The phase factor $e^{i\alpha}$ includes various phases arising in the computation as well as
the frame dependence of $\eta$.
As argued in \cite{HatSuzi}, we will drop this phase assuming that this cancels out with
appropriate prescription for the L\"uscher formula.

The $t$-channel pole at $y^+={X^{(Q)}}^+$ gives exactly the same contribution up to
a phase factor.
Therefore, combining together, we finally obtain the finite-size effect of the two symmetric
DGM configuration as follows:
\be
\delta E_{\mu}=-\frac{64g^2\sin^4\frac{p}{2}}{\sqrt{Q^2+16g^2\sin^2\frac{p}{2}}}
\exp\left[-\frac{2\sin^2\frac{p}{2}\left(L+\sqrt{Q^2+16g^2\sin^2\frac{p}{2}}\right)
\sqrt{Q^2+16g^2\sin^2\frac{p}{2}}}{Q^2+16g^2\sin^4\frac{p}{2}}\right].
\ee
This is exactly what we have derived in Eq.(\ref{stringDGM}) if we identify
$J_1=L,\ J_2=Q$ and $g=\sqrt{\lambda/2}$.

\setcounter{equation}{0}
\section{Concluding Remarks}
In this note we have proposed L\"uscher formula for $\mu$-term
correction of magnon bound states and computed explicitly the
correction for the two symmetric DGMs.
This result is compared with a classical string computation based on Neumann-Rosochatius
reduction.
We showed that the two results match exactly.
This provides another confirmation for the $S$-matrix of the ${\cal N}=6$ CS theory \cite{AN}
in addition to those already investigated \cite{BF,LS}.
It is interesting to apply a similar analysis to asymmetric GM and DGM configurations
on the two $S^2$ spheres.
If the $A$ and $B$ particles are introduced asymmetrically, the $S$-matrix elements
entering into the L\"uscher formula becomes
quite different from those of ${\cal N}=4$ SYM theory.
A similar analysis for ``small GM'' has been performed for one spin case in \cite{BF}
which contains an imaginary value in the correction.
One way of clarifying the unusual result is to do a similar computation for DGMs which
have two spins.
Finally, we emphasize that we have computed only $\mu$-term in this paper
which gives the leading classical limit.
It would be important to extend this result to one-loop order in semi-classical string theory
and compare with the $S$-matrix computation.

\section*{Acknowledgements}
This work was supported in part by KRF-2007-313-C00150 (CA), by
NSFB VU-F-201/06 (PB), and by the Brain Pool program from the
Korean Federation of Science and Technology (2007-1822-1-1).

\end{document}